\documentclass[preprint2]{aastex}\newcommand{\emul}{false}

\newcommand{\IncludeFigures}{true}

\usepackage{amsmath}
\usepackage{amssymb}
\usepackage{xspace}

\usepackage{ifthen}
\newboolean{emul}%
\setboolean{emul}{\emul}%

\ifthenelse{\boolean{emul}}{
  \usepackage{natbib}
  \bibpunct[]{(}{)}{;}{a}{}{,}
  \usepackage{emulateapj}
  \usepackage{apjfonts}
  \setcounter{secnumdepth}{5}
  \usepackage{graphicx}
  \submitted{Draft \today}
}

\lefthead{FRYER, HEGER, LANGER, \& WELLSTEIN}
\righthead{}
\def\gtaprx {\lower .1ex\hbox{\rlap{\raise .6ex\hbox{\hskip .3ex
	{\ifmmode{\scriptscriptstyle >}\else
		{$\scriptscriptstyle >$}\fi}}}
	\kern -.4ex{\ifmmode{\scriptscriptstyle \sim}\else
		{$\scriptscriptstyle\sim$}\fi}}\xspace}
\def\ltaprx {\lower .1ex\hbox{\rlap{\raise .6ex\hbox{\hskip .3ex
	{\ifmmode{\scriptscriptstyle <}\else
		{$\scriptscriptstyle <$}\fi}}}
	\kern -.4ex{\ifmmode{\scriptscriptstyle \sim}\else
		{$\scriptscriptstyle\sim$}\fi}}\xspace}
\def \sun {\ensuremath{_{\scriptscriptstyle \odot}}\xspace}

\newcommand{\K}{{\ensuremath{\mathrm{K}}}\xspace}

\newcommand{\km}{{\ensuremath{\mathrm{km}}}\xspace}
\newcommand{\Msun}{{\ensuremath{\mathrm{M}_{\odot}}}\xspace}
\newcommand{\Sec}{{\ensuremath{\mathrm{s}}}\xspace}
\newcommand{\erg}{{\ensuremath{\mathrm{erg}}}\xspace}

\newcommand{\kms}{{\ensuremath{\km\,\Sec^{-1}}}\xspace}

\newcommand{\kyr}{{\ensuremath{\mathrm{kyr}}}\xspace}

\newcommand{\ms}{{\ensuremath{\mathrm{ms}}}\xspace}

\newcommand{\lSect}[1]{{\label{sec:#1}}}
\newcommand{\lFig}[1]{{\label{fig:#1}}}
\newcommand{\lEq}[1]{{\label{eq:#1}}}
\newcommand{\lTab}[1]{{\label{tab:#1}}}
\newcommand{\pFig}[1]{{\placefigure{fig:#1}}}

\newcommand{\Tabff}[1]{{\ref{tab:#1}}}
\newcommand{\Tab}[1]{{Table~\Tabff{#1}}}

\newcommand{\pan}[1]{{\textit{#1}}}

\newcommand{\FIGFF}[2]{{\ref{fig:#2}\pan{#1}}}
\newcommand{\Figff}[1]{{\FIGFF{}{#1}}}
\newcommand{\FIG}[2]{{Fig.~\FIGFF{#1}{#2}}}
\newcommand{\Fig}[1]{{\FIG{}{#1}}}
\newcommand{\FIGS}[2]{{Figs.~\FIGFF{#1}{#2}}}
\newcommand{\Figs}[1]{{\FIGS{}{#1}}}

\newcommand{\Sectff}[1]{{\ref{sec:#1}}}
\newcommand{\Sect}[1]{{Sect.~\Sectff{#1}}}

\newcommand{\Eqref}[1]{{\ref{eq:#1}}}
\newcommand{\Eqff}[1]{{(\Eqref{#1})}}

\newcommand{\Eq}[1]{{Eq.~\Eqff{#1}}}

\newcommand{\isofont}[1]{{\mathrm{#1}}}
\newcommand{\isomass}[1]{{\ensuremath{\isofont{^{#1}}}}}
\newcommand{\isocharge}[1]{{\ensuremath{\isofont{_{#1}}}}}
\newcommand{\isotope}[3]{{\ensuremath{\isocharge{#1}\isomass{#2}\isofont{#3}}}}

\newcommand{\I}[2]{{\isotope{}{#1}{#2}}}

\newcommand{\Ep}[1]{{\ensuremath{10^{#1}}}}

\newcommand{\Tc}{{\ensuremath{T_{\mathrm{c}}}}\xspace}

\newcommand{\Ye}{{\ensuremath{Y_e}}\xspace}
\newcommand{\fWR}{{\ensuremath{f_{\mathrm{WR}}}}\xspace}
\newcommand{\tWNL}{{\ensuremath{\tau_{\mathrm{WNL}}}}\xspace}
\newcommand{\tWNE}{{\ensuremath{\tau_{\mathrm{WNE}}}}\xspace}
\newcommand{\tWCO}{{\ensuremath{\tau_{\mathrm{WCO}}}}\xspace}
\newcommand{\Cc}{{\ensuremath{\mathrm{C}_{\mathrm{c}}}}\xspace}
\newcommand{\MHe}{{\ensuremath{M_{\mathrm{He}}}}\xspace}
\newcommand{\MCO}{{\ensuremath{M_{\mathrm{CO}}}}\xspace}
\newcommand{\MNeO}{{\ensuremath{M_{\mathrm{NeO}}}}\xspace}
\newcommand{\MSi}{{\ensuremath{M_{\mathrm{Si}}}}\xspace}
\newcommand{\MYe}{{\ensuremath{M_{\mathrm{\Ye}}}}\xspace}
\newcommand{\Ekin}{{\ensuremath{E_{\mathrm{kin}}}}\xspace}

\ifthenelse{\boolean{emul}}{\renewcommand{\pFig}[1]{}}

\newcommand{\FighecoreFile}{fig1.ps} \newcommand{\Fighecore}{
Central carbon abundance (\textsl{upper panel}), stellar mass
(\textsl{lower panel, dotted lines}) and extent of the convective core
(\textsl{lower panel, solid lines}) as a function of the central
helium abundance during central helium burning.  Colors indicate the
different models (see annotation in \textsl{upper panel}).  The total
mass of the star is only displayed after the end of the mass transfer.
The ``kink'' in the evolution of the total mass is caused by the
transition form WNL to WNE, i.e., when the bare helium core is
uncovered.  Note also that the initial central helium abundance of the
helium core is only about 98\,\% at solar metallicity.  The remaining
2\,\% is mostly \I{14}N. \lFig{hecore}}

\newcommand{\FigconvOneFile}{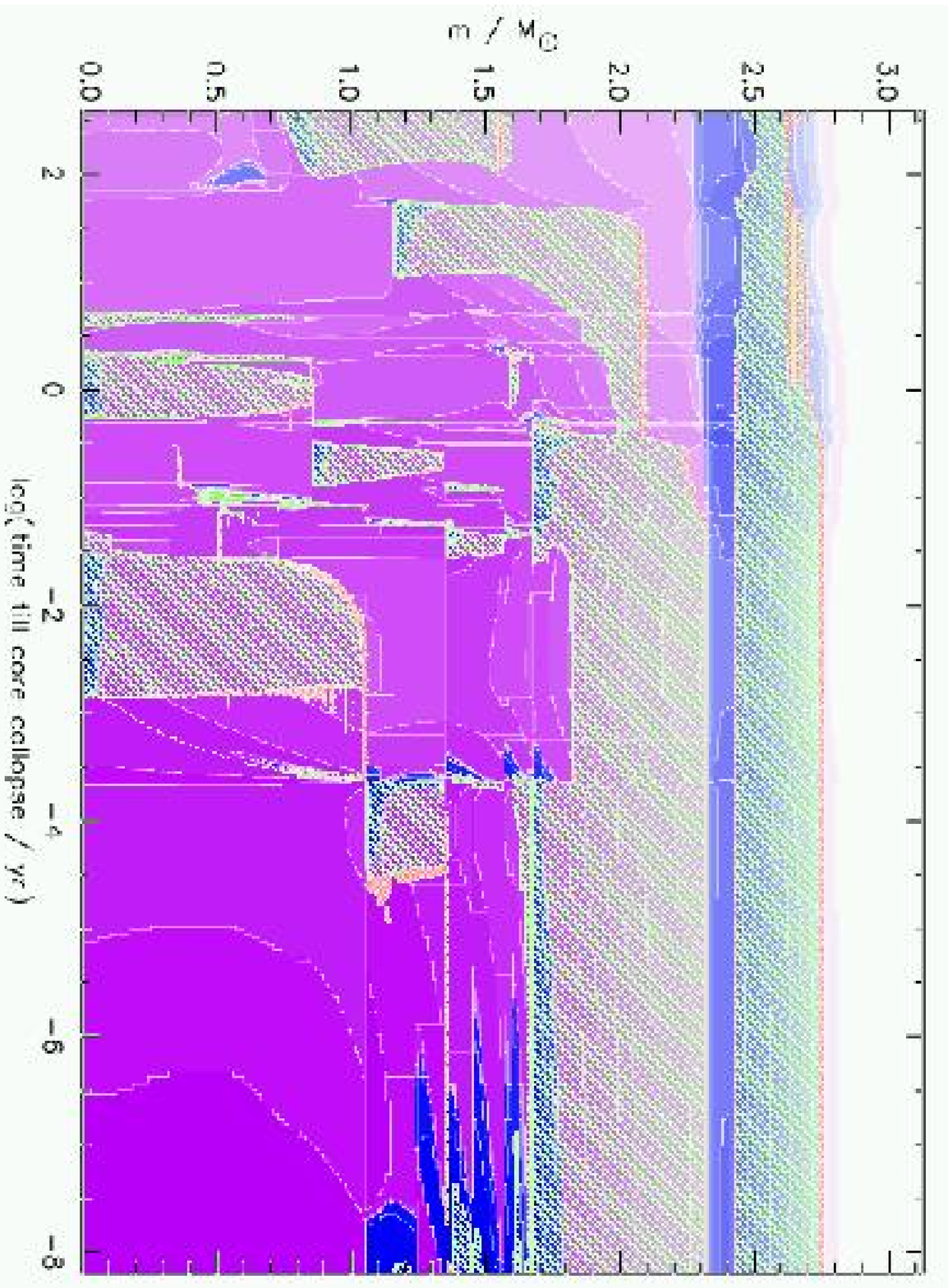} \newcommand{\FigconvOne}{
Kippenhahn diagram of the evolution beyond a central temperature of
$\Tc=\Ep9\,\K$ as followed with the KEPLER code for Model 1s1.  The
\textsl{x-axis} gives the logarithm of the time till core collapse in
years and the \textsl{y-axis} the interior mass coordinate in solar
masses.  \textsl{Green hatching} indicates convective regions and
\textsl{red cross hatching} indicates semiconvective layers.  The
``nuclear'' contribution (burning, photo-disintegration and neutrino
losses) to the energy balance of the star are indicated in
\textsl{blue} (net energy gain) and \textsl{pink} (net energy loss)
shading and the different darker levels code for increases by an order
of magnitude.  The central convection phases shown are neon, oxygen
and silicon burning, followed by their respective shell burning phases
further out.  the preceding series of convective shells are carbon
burning and helium burning occurs in the outermost convective
shell. \lFig{convOne}}

\newcommand{\FigconvTwoFile}{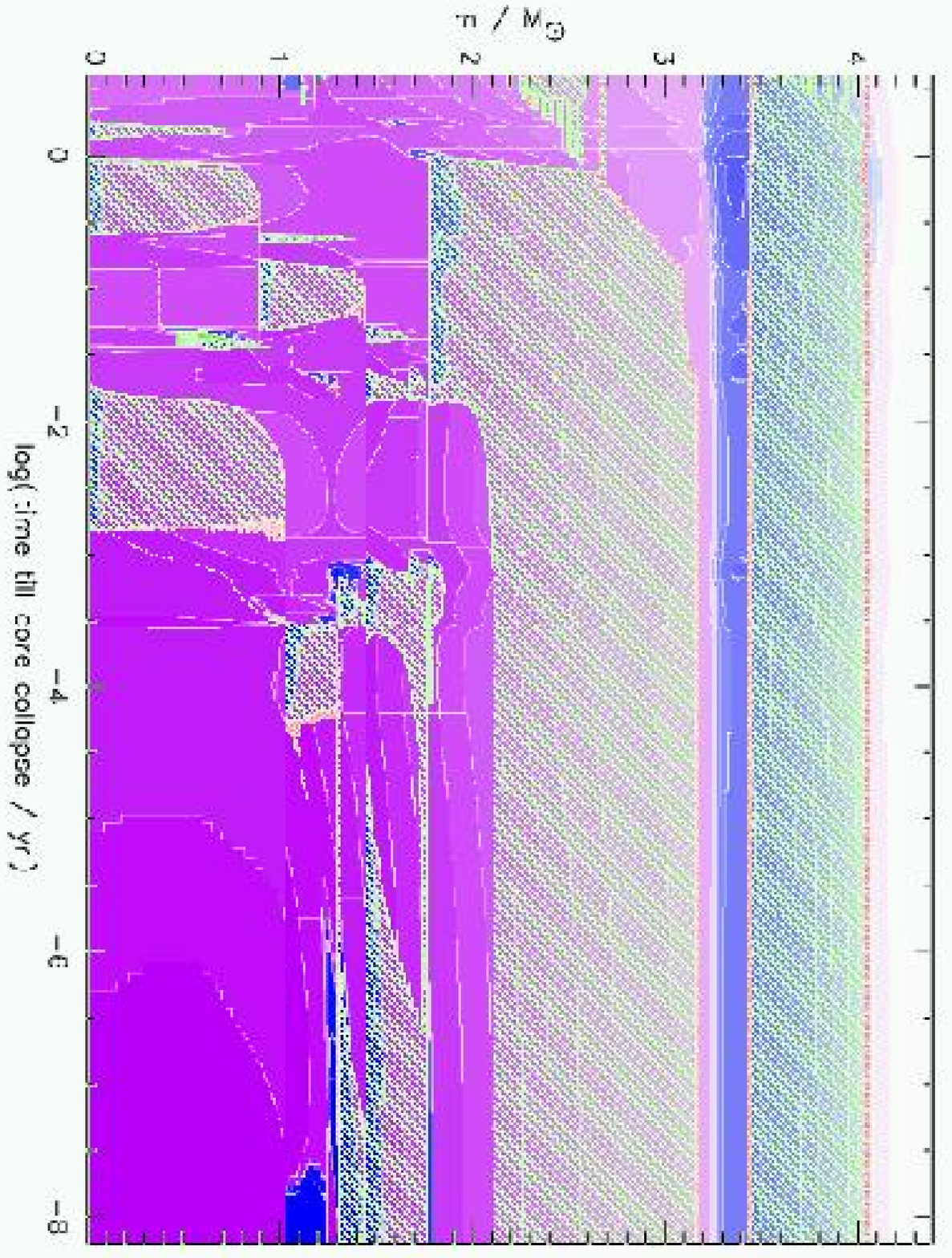} \newcommand{\FigconvTwo}{
Kippenhahn diagram of the evolution beyond a central temperature of
$\Tc=\Ep9\,\K$ as followed with the KEPLER code for Model 1s2.
Colors and shading have the same meaning as in \Fig{convOne}
\lFig{convTwo}} 

\newcommand{\FigconvFourFile}{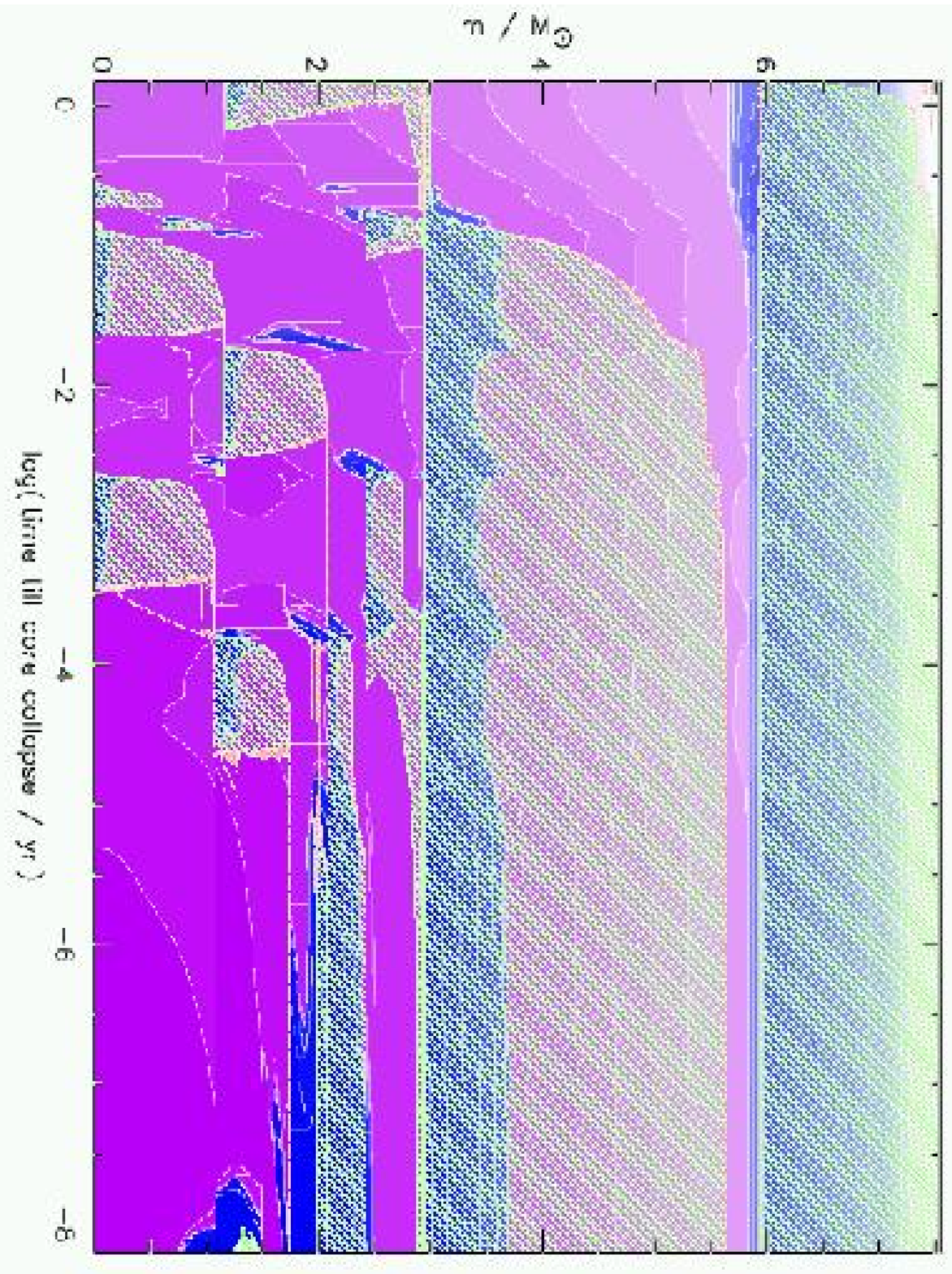} \newcommand{\FigconvFour}{
Kippenhahn diagram of the evolution beyond a central temperature of
$\Tc=\Ep9\,\K$ as followed with the KEPLER code for Model 1s4.
Colors and shading have the same meaning as in \Fig{convOne}
\lFig{convFour}} 

\newcommand{\FigconvSixFile}{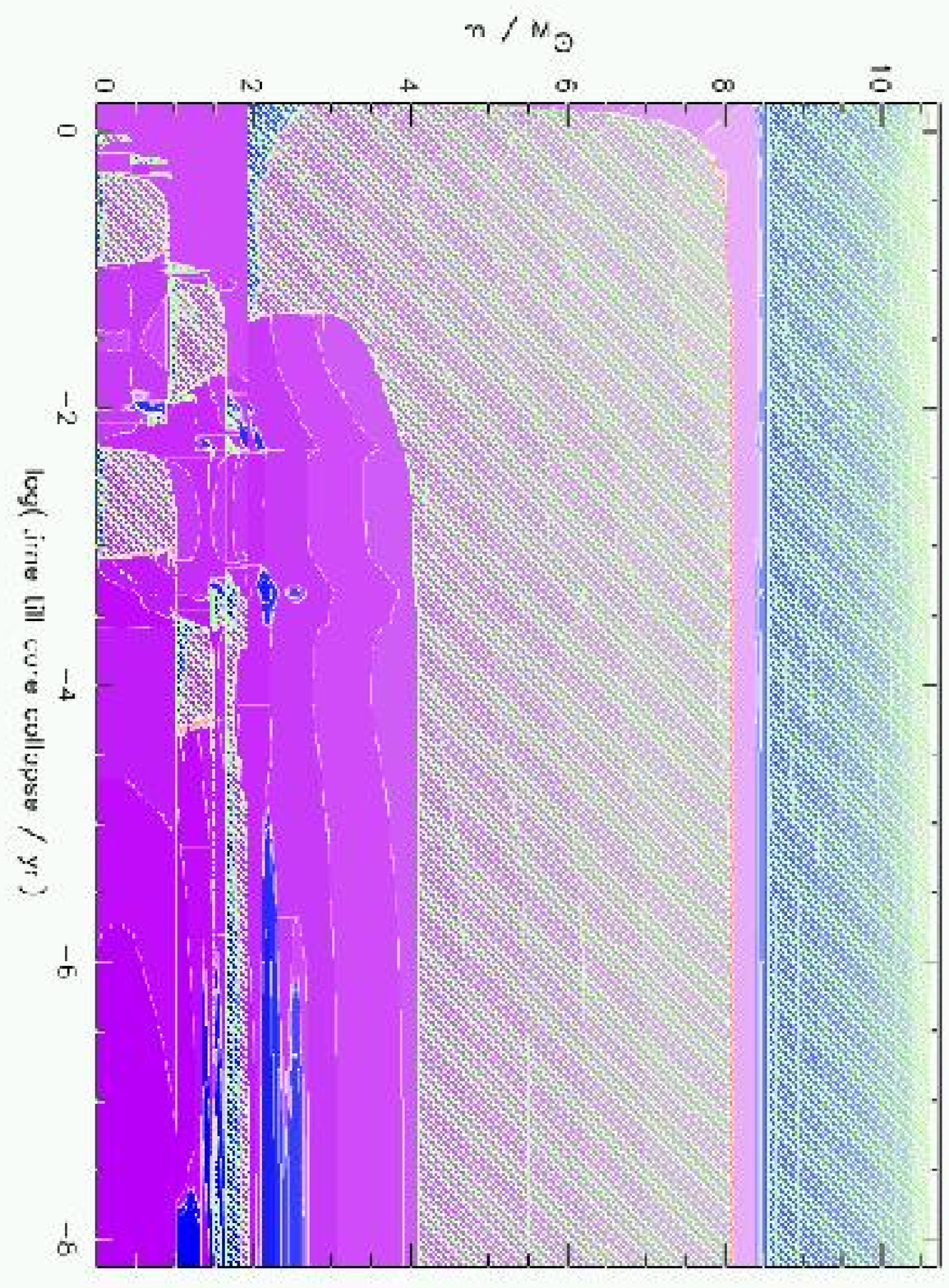} \newcommand{\FigconvSix}{
Kippenhahn diagram of the evolution beyond a central temperature of
$\Tc=\Ep9\,\K$ as followed with the KEPLER code for Model 1s6.
Colors and shading have the same meaning as in \Fig{convOne}
\lFig{convSix}} 

\newcommand{\FigmdotFile}{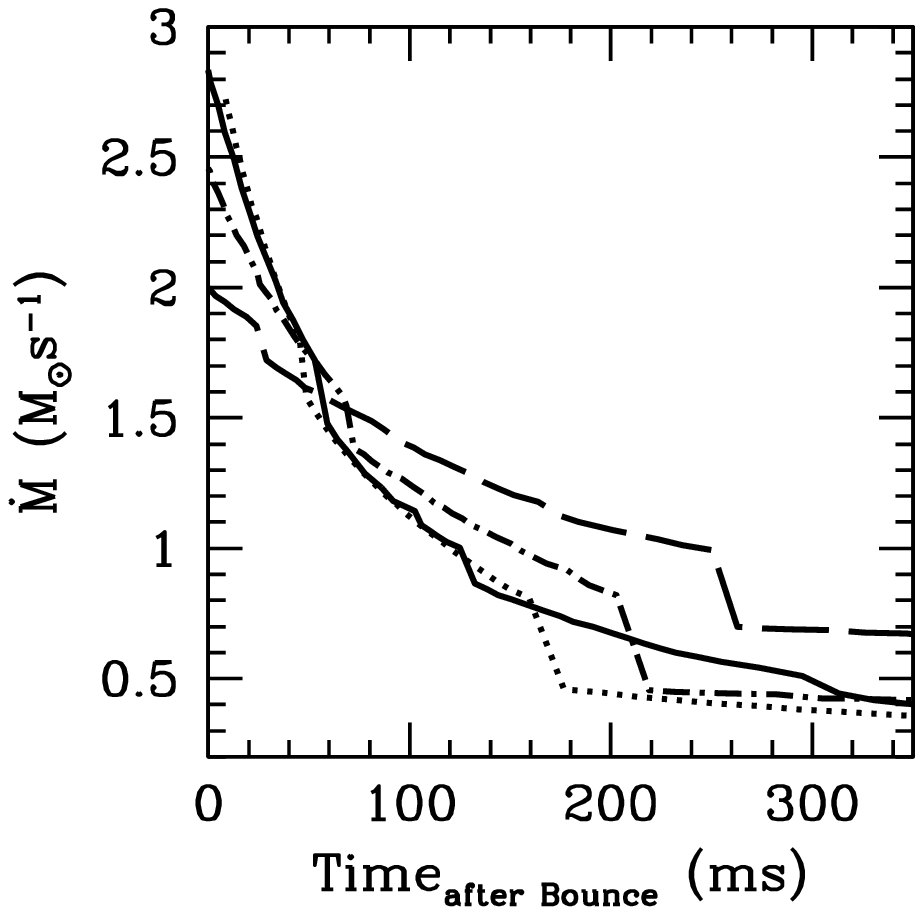} \newcommand{\Figmdot}{Accretion
rate vs.\ time after bounce for Models 1s1 (\textsl{solid}), 1s2
(\textsl{dotted}), 1s4 (\textsl{dashed}), and 1s6
(\textsl{dot-dashed}).  The accretion rates for all of the progenitors
do not differ signifcantly in contrast to the wide range of accretion
rates for 15, 25, and 40\,\Msun progenitors in Fryer (1999).  By
simply following the accretion rate, one might expect that Model 1s1
has the strongest explosion, followed by Model 1s2, Model 1s6, and
finally Model 1s4.  However, our simulations do {\it not} follow this
trend. \lFig{mdot}}

\newcommand{\FigvfFile}{fig7.ps} \newcommand{\Figvf}{Snapshots of the
core-collapse of Models 1s1 (\textsl{upper left}), 1s2
(\textsl{upper right}), 1s4 (\textsl{lower left}), and 1s6
(\textsl{lower right}) 120\,\ms after bounce.
The color denotes radial velocity and the position of the accretion
shock can be easily be determined for each model.  For Models 1s1 and
1s2, the shock is at roughly 650\,km 120\,\ms after bounce.  In
contrast, the shocks of Models 1s4 and 1s6 at the same time are below
500\,km, 400\,km respectively. \lFig{vf}}

\newcommand{\FignuFile}{fig8.ps} \newcommand{\Fignu}{The electron
neutrino luminosity as a function of time after bounce for Models 1s1
(\textsl{solid}), 1s2 (\textsl{dotted}), 1s4 (\textsl{dashed}), and
1s6 (\textsl{dot-dashed}).  Model 1s4, with its large iron core, did
not compress as much during collapse, producing a slightly cooler
core.  A larger fraction of its neutrinos escape as electron neutrinos
(which deposit their energy more efficiently into the convective
region than $\mu$ or $\tau$ neutrinos).  \lFig{nu}}

\newcommand{\FigvsFile}{fig9.ps} \newcommand{\Figvs}{Snapshots of the
core-collapse of the same models as in \Fig{vf} but 170\,\ms after
bounce.  The color denotes radial velocity.  In all the models, the
shock has moved outward over the past 50\,\ms (see \Fig{vf}).  For
Models 1s1 and 1s2, the shock is at roughly 1000\,\km and expanding
rapidly 170\,\ms after bounce.  Model 1s4 also seems to be exploding.
However, the shock of Model 1s6 has stalled at 650\,\km and it appears
no explosion will occur \lFig{vs}}

\newcommand{\FigsfFile}{fig10.ps} \newcommand{\Figsf}{Snapshots of the
core-collapse of four 60\,\Msun progenitors 120\,\ms after bounce.
The color denotes entropy.  This entropy is set by the bounce shock
and it is this entropy profile that seeds the convection in the
convective region.  Note that the entropy is much higher in those 
models with high mass-loss multipliers (1s1 and 1s2). \lFig{sf}}

\newcommand{\FigfallFile}{fig11.ps} \newcommand{\Figfall}{
Mass trajectories more model 1s4.  After 1~d, the mass cut for 
fallback is well defined, but it takes nearly a year for all 
of this material to accrete onto the central black hole. \lFig{fall}}

\slugcomment{Draft for ApJ, \today}

\begin{document}

\title{The Limiting Stellar Initial Mass for Black Hole Formation in 
Close Binary Systems}
\author{C.~L.~Fryer}

\vskip 0.2 in
\affil{Theoretical Astrophysics, MS B288 \\
Los Alamos National Laboratories, Los Alamos, NM 87545}
\authoremail{fryer@lanl.gov}

\author{A.~Heger}
\affil{Department of Astronomy and Astrophysics \\
University of California, Santa Cruz, CA 95064}

\author{N.~Langer}

\affil{Astronomical Institute, P.O. Box 80000,
NL-3508 TA Utrecht, The Netherlands}

\author{and S.~Wellstein}

\affil{Institut f\"ur Physik, Universit\"at Potsdam \\
Am Neuen Palais 10, D-14415 Potsdam, Germany}

\begin{abstract}
We present models for the complete life and death of a 60\,\Msun star
evolving in a close binary system, from the main sequence phase to the
formation of a compact remnant and fallback of supernova debris.
After core hydrogen exhaustion, the star expands, loses most of its
envelope by Roche lobe overflow, and becomes a Wolf-Rayet star.  We
study its post-mass transfer evolution as a function of the 
Wolf-Rayet wind mass loss rate (which is currently not well 
constrained and will probably vary with initial metallicity of the 
star).  Varying this mass loss rate by a
factor 6 leads to stellar masses at collapse that range from
3.1\,\Msun up to 10.7\,\Msun. Due to different carbon abundances left
by core helium burning, and non-monotonic effects of the late shell
burning stages as function of the stellar mass, we find that, although
the iron core masses at collapse are generally larger for stars with
larger final masses, they do not depend monotonically on the final
stellar mass or even the C/O-core mass. We then compute the evolution
of all models through collapse and bounce.  The results range from
strong supernova explosions ($\Ekin > \Ep{51}\,\erg$) for the lower
final masses to the direct collapse of the star into a black hole for
the largest final mass.  Correspondingly, the final remnant masses,
which were computed by following the supernova evolution and fallback
of material for a time scale of about one year, are between 1.2\,\Msun
and 10\,\Msun.  We discuss the remaining
uncertainties of this result and outline the consequences of our
results for the understanding of the progenitor evolution of X-ray
binaries and gamma-ray burst models.

\end{abstract}



\section{Introduction}
\lSect{intro}

It has long been known that stellar-mass black holes could form from
the collapse of massive stars (Oppenheimer \& Snyder 1939) and it is
believed that above some progenitor initial mass limit, stars collapse
to form black holes.  A growing set of evidence suggests that this
mass limit in single stars ($M_{\rm BH}^{\rm S}$) lies somewhere below
25\,M\sun. Fryer (1999) obtained this result from core collapse
simulations.  Maeder (1992) and Kobulnicky \& Skillman (1997) find it
to be most consistent with nucleosynthesis constraints. And the
explosion of a star of about 20\,M\sun as SN~1987A (Arnett et
al. 1989) gives a lower limit to the initial stellar mass required for
direct black hole formation (to be distinguished from black hole
formation due to fall back; see below).  Within the uncertainties of
the above studies, it appears that the black hole mass limit of single
stars is reasonably constrained.

In this paper, we deal with the black hole mass limit for primary
stars of close binary systems ($M_{\rm BH}^{\rm B}$), i.e., for stars
which evolve into the compact objects contained in high and low mass
X-ray binaries or which might become $\gamma$-ray burst sources:
collapsars and black hole binary mergers (see Fryer, Woosley, \&
Hartmann 1999 for a review). As the presence of a companion star only
increases the mass loss of the primary star in a binary system before
the first supernova, the black hole mass limit of binaries is likely
to be larger than that of single stars (i.e., $M_{\rm BH}^{\rm B} >
M_{\rm BH}^{\rm S}$).

There are two observational constraints from observed black hole
systems on the black hole mass limit in binaries.  Using population
synthesis studies, Portegies Zwart, Verbunt, \& Ergma (1997) find that
the number of low mass black hole X-ray binaries in our Galaxy
requires $M_{\rm BH}^{\rm B} < 25\,\Msun$.  Ergma \& van den Heuvel
(1998) argue (without detailed modelling) that the observed periods of
less than $\sim 10\,$h found in most low mass black hole systems are
incompatible with $M_{\rm BH}^{\rm B} > 25\,\Msun$.  While the latter
argument is based on considerations of angular momentum loss
associated with the (uncertain!) Wolf-Rayet winds.  In both
investigations neither $M_{\rm BH}^{\rm S}$ and $M_{\rm BH}^{\rm B}$
are distinguished, nor do they consider the dependence of $M_{\rm
BH}^{\rm B}$ on the type of binary evolution.

Wellstein \& Langer (1999; WL99) showed that indeed $M_{\rm BH}^{\rm
B}$ may be very different from $M_{\rm BH}^{\rm S}$, and is
strongly dependent on the type of binary interaction.  Only in the so
called Case~C systems --- i.e., in initially wide systems where mass
transfer starts only after the primary has evolved through the major
part of core helium burning --- may the black hole mass limits be
comparable; i.e., $M_{\rm BH}^{\rm BC} \simeq M_{\rm BH}^{\rm S}$
(Brown, Lee, \& Bethe 1999).  However, due to the Wolf-Rayet winds,
which reduce the total stellar mass and thus the helium core mass {\em
during} core helium burning, the black hole mass limit for Case~A and
Case~B binaries (mass transfer starts during or directly after core
hydrogen burning), is clearly smaller; i.e., $M_{\rm BH}^{\rm BA} >
M_{\rm BH}^{\rm BC}$ and $M_{\rm BH}^{\rm BB} > M_{\rm BH}^{\rm BC}$.
And although WL99 found that $M_{\rm BH}^{\rm BA} \simeq M_{\rm
BH}^{\rm BB}$, this result is expected to be different (i.e., $M_{\rm
BH}^{\rm BA} > M_{\rm BH}^{\rm BB}$) if less efficient Wolf-Rayet wind
mass loss were assumed (i.e., compare our Model 1s2 below with Model
2$^{\prime}$ of WL99).

The relevance of detailed binary evolution models for the significance
of constraints on $M_{\rm BH}^{\rm B}$ from observed X-ray binaries
has been demonstrated by WL99.  Ergma \& van den Heuvel (1998) argued
that the pulsar GX301-2, which has a 40...50\,M\sun supergiant
companion, originated from a star of more than 50\,M\sun, implying
that $M_{\rm BH}^{\rm B} > 50 M_{\odot}$ for this particular system.
However, WL99 computed detailed Case~A progenitor evolution models
which satisfy all observational constraints for this system, but in
which the pulsar progenitor has an initial mass of only
26\,M\sun. This solution implies $M_{\rm BH}^{\rm S} < M_{\rm BH}^{\rm
BA} < 26 M_{\odot}$.

In the following we investigate in detail the black hole mass limit
for primary stars in close binary systems, using a particular Case~B
evolutionary sequence with 60\,M\sun primary star. We study five cases
where we only change the stellar wind mass loss rate in the Wolf-Rayet
phase of the primary star which we vary over the anticipated regime of
uncertainty, i.e., by a factor of 6.  For high mass loss rates (e.g.,
Hamann, Koesterke \& Wessolowski 1995, Langer 1989b, Braun 1997,
Woosley, Langer, \& Weaver 1995), the stellar mass at collapse can
reduce down to $\sim$3\,M\sun with an iron core mass as low as
1.3\,M\sun (Woosley, Langer, \& Weaver 1995).  The lower Wolf-Rayet
mass loss rates proposed by Hamann \& Koesterke (1998) and Nugis \&
Lamers (2000) result in higher final masses (WL99).

For all five cases, we model the evolution of the primary star up to
iron core collapse (Sect.~2), which predicts the final stellar
mass and the detailed pre-supernova structure of these objects ---
e.g., their final iron core mass. We then use these pre-collapse
structures to model their collapse through bounce and explosion (if
applicable; Sect.~3). Finally, we follow the evolution of the
supernova ejecta for about one year, which allows us to
estimate the final remnant mass including fall back of
previously ejected material onto the compact object. With these
coupled simulations we obtain the dependence of the compact remnant
mass in a given binary system on the assumed strength of the
Wolf-Rayet wind.  Implications of our results for determinations of
the black hole mass limit and the progenitor evolution of X-ray
binaries are discussed in Sect.~4.

\section{Pre-Supernova Evolution}
\lSect{preSN}

In this section, we describe the evolution of the primary star
from the end of the Case~B mass transfer until core collapse. 
The earlier evolutionary stages are described in WL99.

\subsection{Assumptions and method}

Based on evolutionary models of Wolf-Rayet stars, and by comparing
observed number ratios of Wolf-Rayet to O~stars and carbon to nitrogen
rich Wolf-Rayet stars in our Galaxy, Langer (1989b) has proposed a
mass-dependent mass loss rate for Wolf-Rayet stars. At the time, this
mass loss rate was consistent with results derived from Wolf-Rayet
wind model atmospheres for individual objects (e.g., Hamann et al.\
1995).  With this rate, Woosley et al.\ (1995) studied the evolution
and pre-supernova structure of helium stars.  They found fairly small
final stellar masses, even for the initially most massive stars in
their sample.  The inclusion of ``clumping'' in Wolf-Rayet atmosphere
models by Hamann \& Koersterke (1998) recently suggests that the mass
loss rates of these stars could be considerably smaller.  Hamann \&
Koersterke proposed a clumping ratio of 2$\ldots$3, i.e., a reduction
of the mass loss rate by the same factors.  This agrees with the
recent, independant Wolf-Rayet mass loss rate determinations of Nugis
\& Lamers (2000).

Binary star models with reduced Wolf-Rayet mass loss rates have
already been presented by WL99.  Here we concentrate on their most
massive Model~1 (see Table~1 of WL99), i.e., a binary system
consisting of a 60\,\Msun and a 34\,\Msun companion with an initial
orbital period of 20\,days.  Note that the primary's evolution is
rather independant of the secondary mass and of the orbital period, as
long as the system remains a Case~B system, and as long as a merging
of both binary components is avoided.  We also consider WL99's Models
1$^{\prime}$, and 1$^{\prime\prime}$ which use 1/2, and 1/4 times
their ``standard'' Wolf-Rayet mass loss rate --- which is similar to
that of Langer (1989b).  Here, we denote these respectively as Models
1s1, 1s2, and 1s4.  We computed additional models with 1/3 (Model 1s3)
and 1/6 (Model 1s6) times their standard Wolf-Rayet mass loss rate.
This may cover the range of uncertainty, and it allows us to explore
in detail the implications for the pre-supernova structure and the
supernova explosions.


The evolution of the binary system prior to and through the mass
transfer, and through the Wolf-Rayet mass loss phase was computed with
the stellar evolution code described by WL99 (see also Langer 1991b;
Braun 1997; Langer 1998).  Before ignition of central neon burning, at
a central temperature of \Ep9\,\K, the calculation is stopped, and the
further evolution is followed using the KEPLER code (Weaver,
Zimmerman, \& Woosely 1978), similar to the calculations of Woosley,
Langer, \& Weaver (1993), and that of rotating stars by Heger, Langer,
\& Woosley (2000).  Since both codes use a very similar equation of
state, this ``link'' was unproblematic and did not result in noticeable
adjustments of the stellar structure.  We followed the evolution of
the primary star as a single star until ``onset of core collapse'',
which we define as the time when the infall velocity in the core
exceeds 900\,\kms (Woosley \& Weaver 1995).  At this point, the
one-dimensional stellar evolution calculation is stopped.  The
subsequent core collapse calculations and the start of the supernova
explosion are discussed in \S3.

\subsection{The Wolf-Rayet phase}

\pFig{he-kern}
\ifthenelse{\boolean{emul}}{
\vspace{1.5\baselineskip}
\noindent
\includegraphics[angle=0,width=\columnwidth]{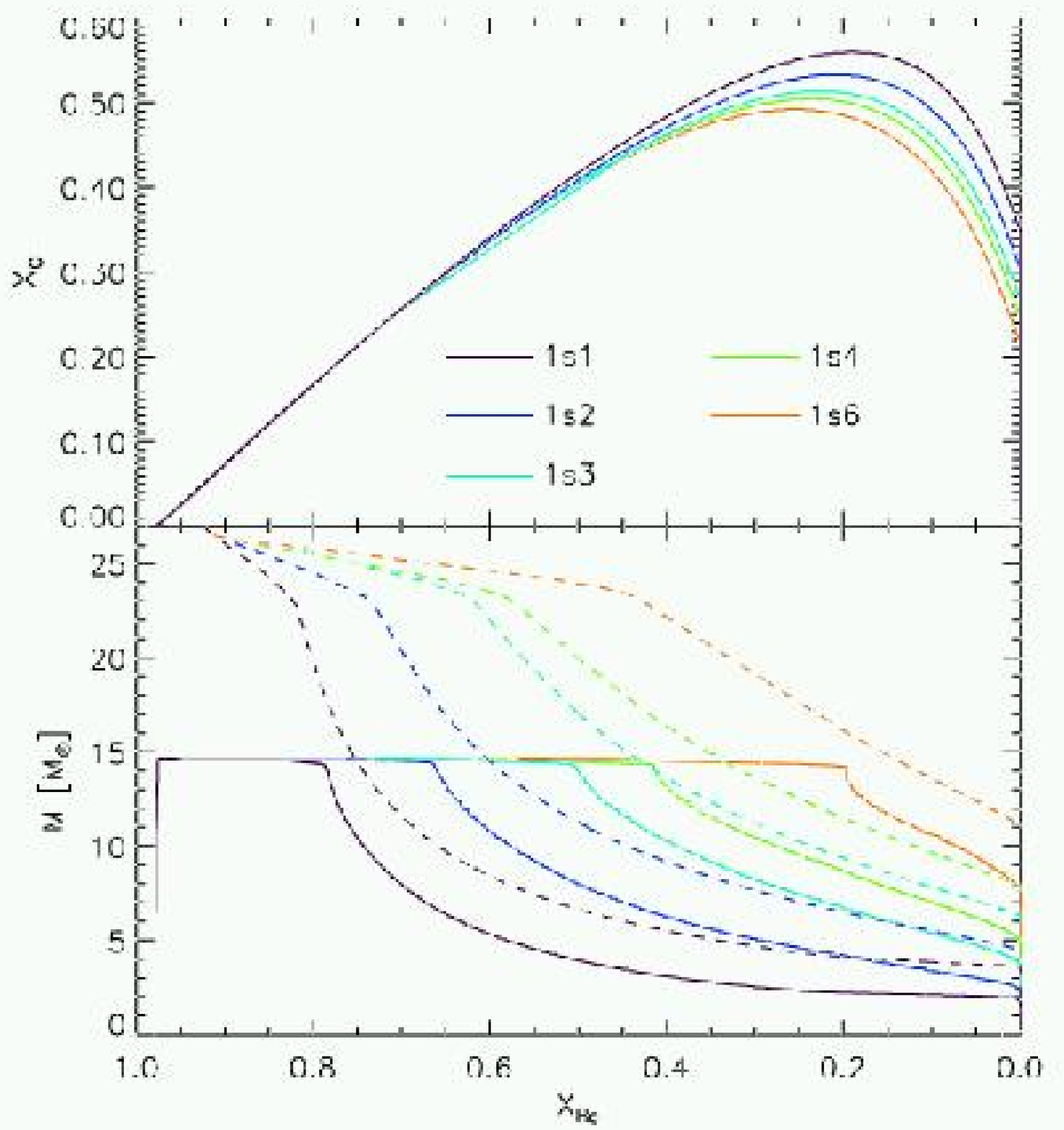}
\figcaption{\Fighecore}
\vspace{1.5\baselineskip}}{}

\pFig{convOne}
\ifthenelse{\boolean{emul}}{
\vspace{1.5\baselineskip}
\noindent
\includegraphics[angle=0,width=\columnwidth]{\FigconvOne}
\figcaption{\FigconvOne}
\vspace{1.5\baselineskip}}{}

\pFig{convTwo}
\ifthenelse{\boolean{emul}}{
\vspace{1.5\baselineskip}
\noindent
\includegraphics[angle=0,width=\columnwidth]{\FigconvTwo}
\figcaption{\FigconvTwo}
\vspace{1.5\baselineskip}}{}

\pFig{convFour}
\ifthenelse{\boolean{emul}}{
\vspace{1.5\baselineskip}
\noindent
\includegraphics[angle=0,width=\columnwidth]{\FigconvFour}
\figcaption{\FigconvFour}
\vspace{1.5\baselineskip}}{}

\pFig{convSix}
\ifthenelse{\boolean{emul}}{
\vspace{1.5\baselineskip}
\noindent
\includegraphics[angle=0,width=\columnwidth]{\FigconvSix}
\figcaption{\FigconvSix}
\vspace{1.5\baselineskip}}{}

The massive primary star undergoes two major mass-loss phases.
First, it loses mass during a mass-transfer phase due to Roche lobe
overflow, which almost uncovers its helium core (see
below). Then, Wolf-Rayet winds begin to dominate the mass-loss.  
Here we discuss how the evolution of the primary varies
for different Wolf-Rayet mass-loss multipliers. 

The 60\,\Msun primary star loses most of its hydrogen envelope during
the mass transfer phase which ends shortly before the onset of central helium
burning, reducing the primary mass to 26.8\,\Msun.  Then the
star experiences a Wolf-Rayet stage in which hydrogen is
still present (the ``WNL'' phase).  
This stage lasts between 25 and 122\,\kyr (see
\Tab{evo}) and ends when the hydrogen-rich layers have been lost
completely.  The luminosity of the star depends primarily on the size
of the hydrogen-free core, which is roughly the same for all of our
models during this phase.  Since, in our
simulations, the mass loss depends only on the luminosity and surface
hydrogen abundance of the star (see equation 1 of WL99),
the total mass lost during this phase scales almost directly with the
Wolf-Rayet mass-loss multiplier.  Hence, the lifetime of this phase
roughly scales inversely with this multiplier.  However, the models
with lower mass loss rates eventually become slightly more luminous
with time since the helium core evolves and grows slightly due to
hydrogen shell burning.  So the lifetime does not {\em strictly} scale
with the inverse of the mass loss rate but is slightly
shorter than the inverse scaling law imples (Table 1).

When the helium core is completely uncovered, the star 
passes through a ``WNE'' phase and the mass loss
rate increases --- this corresponds to the ``kink'' in the line for
the total mass of the star in the lower panel of \Fig{hecore}, at a
central helium abundance of about 82\,\% in Model 1s1 and 45\,\% in
Model 1s6.  It ends when the star uncovers the layers that are
enriched in products of central helium burning.  This occurs when the
mass drops below the maximum extent in mass of the convective core, or
--- in particular in the models with smaller mass loss rates --- in
the partially mixed layers above the convective core (see Fig.~1 of
Langer 1991a).  This marks the end of the WNE and the beginning of 
the WC Wolf-Rayet phase.

The total mass lost due to the Wolf-Rayet wind depends much weaker
than linearly on the mass loss multiplier (e.g. Model 1s1 loses only
1.5 times as much mass as Model 1s6 through Wolf-Rayet winds; 
Table~1).  The reason is that, unlike in the WNL phase, the helium
core mass and its chemical composition change considerably during the
WNE phase.  Since the stellar luminosity depends on both quantities,
this feeds back to the mass loss rate.  Assuming the total pressure in
the star is due to equal contributions from ideal gas and radiation
pressure --- which is the case in a Wolf-Rayet star of about 20$\,
M_{\odot}$ --- one obtains from homology considerations for chemically
homogeneous stars (Kippenhahn \& Weigert 1990) with constant opacity
\begin{equation}
L\propto  M^{\alpha} \mu^{\beta}~   ,
\end{equation}
with $\alpha=1.4$ and $\beta= 2...3$.  Thus, for constant chemical
composition the mass loss rate depends only on the stellar mass. In
this case, the mass loss rate of our models were only a function of
the stellar mass.  However, the mean molecular weight, $\mu$, changes
in the course of helium burning from $\mu \simeq 1.33$ in the beginig
to $\mu \simeq 1.8$ at the end. According to the above relation, this
might change the luminosity of a Wolf-Rayet model of fixed mass by a
factor of $\sim2$.  Even though in more realistic models this effect
is smaller (Langer 1989a), we see that their mass evolution does not
follow a simple power law, and that the total amout of mass lost during
the Wolf-Rayet stage is not a simple function of the mass loss
multiplier.
For example, models computed with a lower Wolf-Rayet mass-loss rate
enter the WNE phase roughly with the same mass as those with higher
mass loss rate, but they do so later in time, i.e., they have by then a
chemically more evolved core (\Fig{hecore}).  Therefore, they have
an increased luminosity at that time, which results in a higher mass
loss rate than simply scaling it with the mass loss
multiplier would predict.

Additional effects can influence the mass loss rate, e.g., the feature
that the convective cores of chemically evolved stars recede later 
(\Fig{hecore}). Thus, even if the luminosity is the only physical
parameter on which the the WNE mass loss rate depends, it varies in
non-trivial ways as function of time.

When the products of helium burning appear at the surface of the star,
the star enters a ``WC'' phase.  By then, the stellar mass has
decreased significantly compared to earlier evolution stages and the
star's evolutionary time-scale becomes much longer than those in the
WNL or WNE phases (see Table~1).  Only for the \emph{lowest} mass-loss
rate multiplier (Model 1s6; biggest presupernova stellar masses) does
the surface oxygen mass fraction become larger than the carbon mass
fraction.  The final surface helium mass fraction of this model is
only 10\,\%, while it is 40\,\% in Model 1s1.  The final masses are in
the range 3.1\,\Msun$\ldots$10.7\,\Msun (\Tab{evo}).


\subsection{From carbon burning to core collapse}
\lSect{Cc2cc}

An important consequence of the different evolution of the helium core
mass (\Fig{hecore}) is different core carbon abundances at central
helium depletion (\Tab{evo}).  This and the C/O-core mass determine
the duration and extent of the carbon burning phases and of all later
burning stages.  The central carbon abundances at core helium
exhaustion, \Cc, is larger for larger mass loss from the helium core
during core helium burning.  That is, it is much larger in the primary
components of close binary systems than in single stars of the same
initial mass (WL99; Brown et al.\ 2001).  And for the same reason, it
is a monotonic function of the mass loss multiplier in our study.
We obtain the largest value in Model~1s1 (Table~1).

After central helium exhaustion, Models 1s1$\ldots$1s3 experience
convective core carbon burning, while in Models 1s4 and 1s6 the
central carbon abundance is too low and the central burning phase
remains radiative.

Core carbon burning is followed by carbon shell burning. In Models
1s2$\dots$1s4 the convective carbon shell burns through its fuel and
extinguishes before neon ignition (\Figs{convOne} -- \Figff{convFour};
Model 1s1 has two phases of carbon shell burning before core neon
ignition). The upper edge of the burning shell then marks the edge of
the ``carbon-free'' core which is bigger in more massive models in our
simulations.
Thus we find that lower mass loss rates produce bigger
carbon-free cores.  Correspondingly, the silicon and iron core masses
increase for lower mass loss multipliers in our model sequence.

However, Model 1s6 shows that this trend is not universal
(\Fig{convSix}).  In this model, even most of the neon and oxygen burning
proceeds while the first carbon burning shell, which is located at only
1.9\,\Msun, is still active.  Hence, the carbon-free core is limited
to the region below this shell until this late time.  For cores of this high
mass, the remaining evolution time-scale is too short for the former
carbon shell to become a fully integrated part of the Ne/O core and
some traces of carbon remain even until core collapse.  Oxygen burning
remains restricted to the region below, limiting the size of the
silicon core to this mass.  This is in contrast to, e.g., Model 1s4
where a much larger carbon-free core is established already before
neon ignition, resulting in bigger silicon and iron cores.  Therefore
Model 1s6 ends up with a {\em smaller} iron core than Models 1s4 and
1s3, despite the fact that it has by far the largest total mass
(Table~1).

A similar non-monotonic behaviour can occur at the transitions from
two to three shells, and so forth.  In our case, the reason is the
change in the mass loss rate, but in general they depend on the
core mass,
the carbon abundance in the CO-core (Boyes 2001), and the mass and
composition of the overlaying helium shell.  Clearly, further studies
are required to understand this phenomenon in more detail (Boyes,
Heger, \& Woosley 2001).

Models~1s1 and~1s2 end their lives as rather small helium cores
(\Tab{param}).  Their evolution is influenced by partial electron
degeneracy and results in the formation of rather small iron cores
(\Tab{60msun}), i.e., of the order their Chandrasekhar mass, after one
phase of silicon shell burning (\Figs{convOne} and \Figff{convTwo}).
In Models~1s3 and~1s4 (\Fig{convFour}) the silicon shell reaches out
to higher masses, resulting in bigger iron cores. This is a
consequence of the larger carbon-free core established before
beginning of central neon/oxygen burning.  In Model 1s6, a small
carbon-free core of 1.9\,\Msun is maintained until very late,
i.e., until the first oxygen shell burning is extinguished.  (\Fig{convFour};
note the logarithmic time axis).  Only then does the carbon shell
extinguish, so that only the region below this shell participates in the
subsequent burning stages.  This leads to a similar late time
evolution and final iron core size less massive than Models~1s3 and~1s4.  
However, as we shall see in \S 3, the mass of the deleptonized 
core is not the only factor governing the remnant mass after 
collapse, and even though Model 1s6 has a less massive iron 
core, it produces a larger compact remnant.

There are only few calculations in the literature that allow a
comparison with our study.  Woosley, Langer, \& Weaver (1995) modeled
the evolution of helium stars using the Wolf-Rayet mass loss rates of
Langer (1989).  Their most massive model had an initial mass of
20\,\Msun, which is similar to the initial helium core mass of our
models ($\sim23\,\Msun$).  Their resulting final stellar mass
(3.55\,\Msun), CO core mass (2.53\,\Msun) and NeO core mass
($\sim1.8\Msun$) lie between that of our Models~1s1 and~1s2, but the
resulting iron core mass is bigger (1.49\,\Msun),
possibly due to interactions of oxygen and silicon burning shells 
that lead to a peculiar behavior of our Model 1s2 (see \Fig{convTwo}). 

\section{Core-Collapse, Supernovae, and Compact Remnants}

\subsection{Numerics}

We follow the core collapse of the final Wolf-Rayet star models
described above using the 2-dimensional, smooth particle
hydrodynamics, core-collapse supernova code originally described in
Herant et al.\ (1994).  This code follows the core collapse
continuously from collapse through bounce and ultimately to explosion.
The code includes a variety of neutrino rates and cross-sections to
model 3 neutrino populations ($\nu_e, \bar{\nu}_e$, and
$\nu_\mu+\nu_\tau$) and transports these neutrinos via a flux-limited
diffusion algorithm.  The equation of state is a patchwork of a series
of codes valid over the range of densities and temperatures required
in the course of the core-collapse simulation (see Herant et al.\ 1994
for details).  In the regime of low density, the equation 
of state is identical to that of KEPLER.  Hence, we do 
not encounter problems ``matching'' the output of KEPLER into the 
core-collapse code.  To this code, Fryer et al.\ (1999) added
spherically symmetric general relativity and a more sophisticated flux
limiter.

In all our simulations, we model the inner $\sim 4.3$\,\Msun of the
star (for Model 1s1 we model the entire 3.1\,M\sun star) with
13,000-16,000 particles in a 180$^\circ$ hemisphere.  We assume
rotational symmetry about the hemisphere causing each particle to
effectively represent a ring about the axis of symmetry.  The
advantage of this 2-dimensional, hydrodynamics code is that it allows
us to model from collapse through explosion with reasonable angular
resolution ($\sim 1^{\circ}$) without requiring us to reset the grid
at any point in the simulation.  In addition, we need only remove the
inner 0.001\,\Msun which minimizes the effects of the inner boundary.

This code has been used for a variety of core-collapse
simulations (Fryer et al.\ 1999, Fryer 1999, Fryer \& Heger 2000)
which provide a basis with which to compare our models.  By comparing
the collapse results of the 60\,\Msun cores in this paper with
themselves, and with the core-collapse results of other massive
progenitors (Fryer et al.\ 1999, Fryer 1999), we can determine trends
caused by differences in the progenitors alone (and not the
core-collapse code).

\subsection{Core-Collapse and SN Explosions:  Cause and Effect}

Table 2 summarizes the results of our core-collapse simulations.  Note
that for decreasing Wolf-Rayet mass-loss rate, the explosion energy
initially grows (Model 1s1 vs. Model 1s2).  But as we continue to
decrease the mass-loss rate, the explosion energy decreases down to
Model 1s6 (which does not explode at all).  This initial rise, and
then decrease, in explosion energy mimics the trend in explosion
energy of increasing progenitor mass without mass-loss (Fryer et al.\
1999, Fryer 1999).  The decrease in explosion energy with increasing
progenitor mass fits easily into a simple picture of the
neutrino-driven engine in which a convective region must overcome the
ram pressure of infalling material in order to launch an explosion
(see, for example, Burrows \& Goshy 1993).  The ram pressure at the
accretion shock is given by
\begin{equation}\lEq{pshock}
P_{\rm shock}=\dot{M}_{\rm shock} \sqrt{2 G M_{\rm enclosed}}/
(8 \pi r_{\rm shock}^{2.5}),
\end{equation}
where $\dot{M}_{\rm shock}$ is the accretion onto the convective
region, $G$ is the gravitational constant, $M_{\rm enclosed}$ is
the enclosed mass below the accretion shock and $r_{\rm shock}$
is the radius of the shock which caps the convective region.

More massive progenitors produce stars which have higher accretion
rates onto the accretion region.  Fryer (1999) found that these more
massive stars could not explode until the accretion rate onto the
convective region dropped significantly.  Hence, more massive stars
explode later.  Their neutrino luminosity (and subsequent heating) at
these later explosion times tend to be lower so that their explosions
are weaker.  Extremely low-mass progenitors (accretion induced
collapse of white dwarfs, progenitors between 8-11\,\Msun) have very
little infalling material, and the explosion occurs before significant
neutrino-energy deposition.  Hence, these low-mass progenitors also
have explosion energies which are weaker than those of stars within
the $\sim 13-20$\,\Msun regime.

The pressure of the accreting material depends both upon the accretion
rate of the infalling material and the position of the shock (which
marks the boundary between the convective region and the collapsing
star).  For a given accretion rate, the ram pressure of the infalling 
material decreases as the shock progresses further out from the star 
(see, for example, Burrows \& Goshy 1993, Janka 2000).  By
combining the accretion rates from our simulations (Fig. 6) with the
shock radii 120\,ms after bounce (Fig. 7), we can estimate the
pressure that the convective region must overcome to launch an
explosion.  Fig. 7 shows in color coding the radial velocity of Models
1s1, 1s2, 1s4, and 1s6 at 120\,ms after bounce.  The position of the
accretion shock can be easily be determined as the interface of low
velocity (white) at the bottom of rapid inward movement (red).  For
Models 1s1 and 1s2, the shock is at roughly 650\,km.  In contrast, in
Models 1s4 and 1s6 the shock is below 500\,km and 400\,km,
respectively.  Using \Eq{pshock}, the corresponding shock pressures
for Models 1s1, 1s2, 1s4, and 1s6 are 4.6, 4.5, 13.2, and 18.4
$\times 10^{25}$\,erg\,cm$^{-3}$, respectively .

\pFig{vf}
\ifthenelse{\boolean{emul}}{
\vspace{1.5\baselineskip}
\noindent
\includegraphics[angle=0,width=\columnwidth]{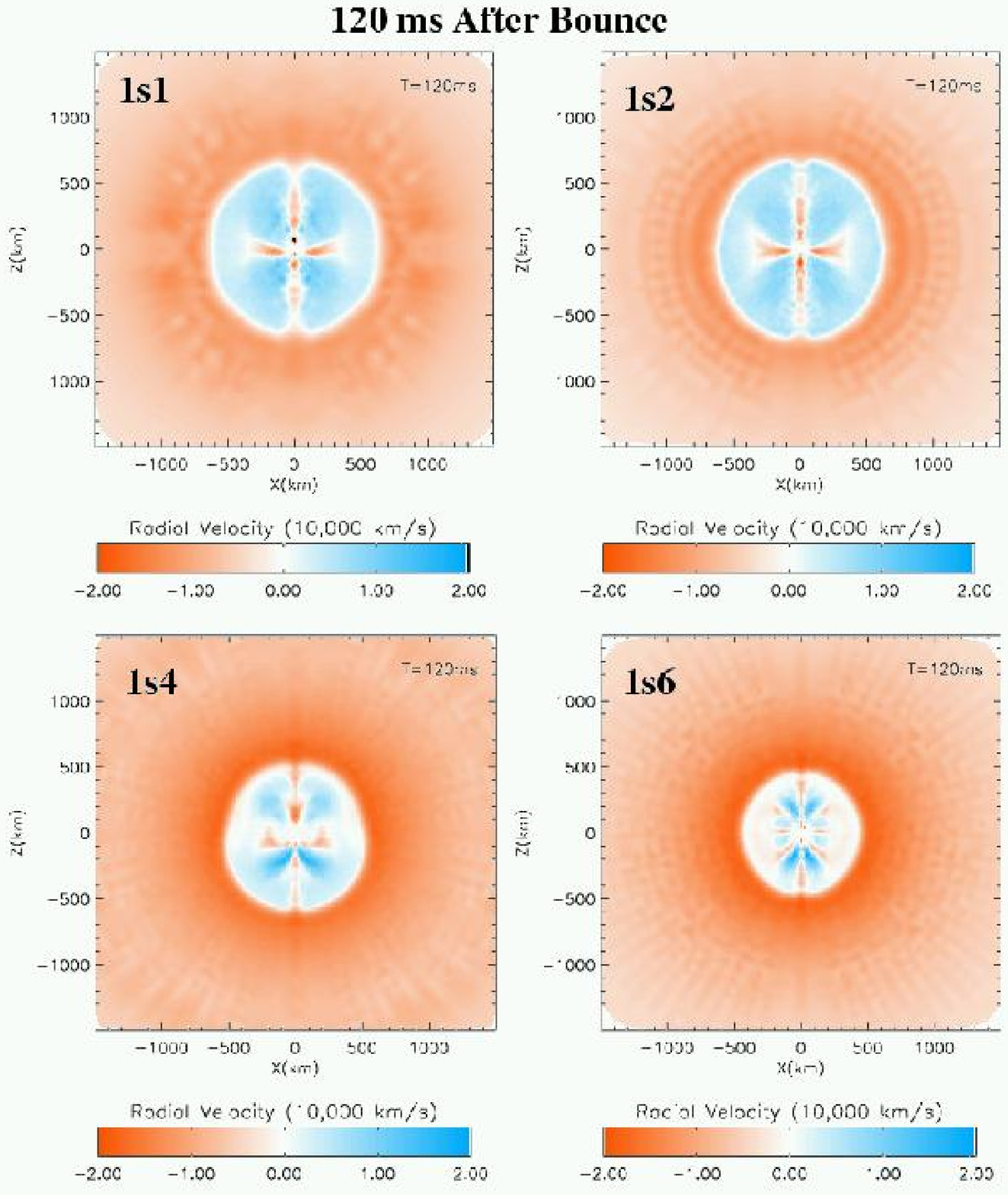}
\figcaption{\Figvf}
\vspace{1.5\baselineskip}}{}

170\,ms after bounce, we can already see the differences in the shock
pressure manifesting themselves in the explosion evolution (Fig.\ 8).
Models 1s1 and 1s2 have strong explosions with rapidly outward-moving
shock radii at 1000\,km, while Model 1s4 is exploding weakly (the
shock is at 800\,km), and Model 1s6 reaches its peak at 650\,km.  The
trend of explosion energies matches the initial trend in the shock
pressures just after bounce (Table 2).  By calculating the pressure at
the shock shortly after the shock stalls from the shock position and
infall accretion rate, we can gain insight into the fate of the
collapsed star (i.e., whether it will become a black hole or neutron
star).  The shock pressure is clearly a better diagnostic of resultant
supernova explosion energy than the accretion rate alone, but,
unfortunately, it can not be determined unless one models the collapse
(at least through bounce) of the massive star.  Fortunately, such
calculations can be done in 1-dimension.

\pFig{vs}
\ifthenelse{\boolean{emul}}{
\vspace{1.5\baselineskip}
\noindent
\includegraphics[angle=0,width=\columnwidth]{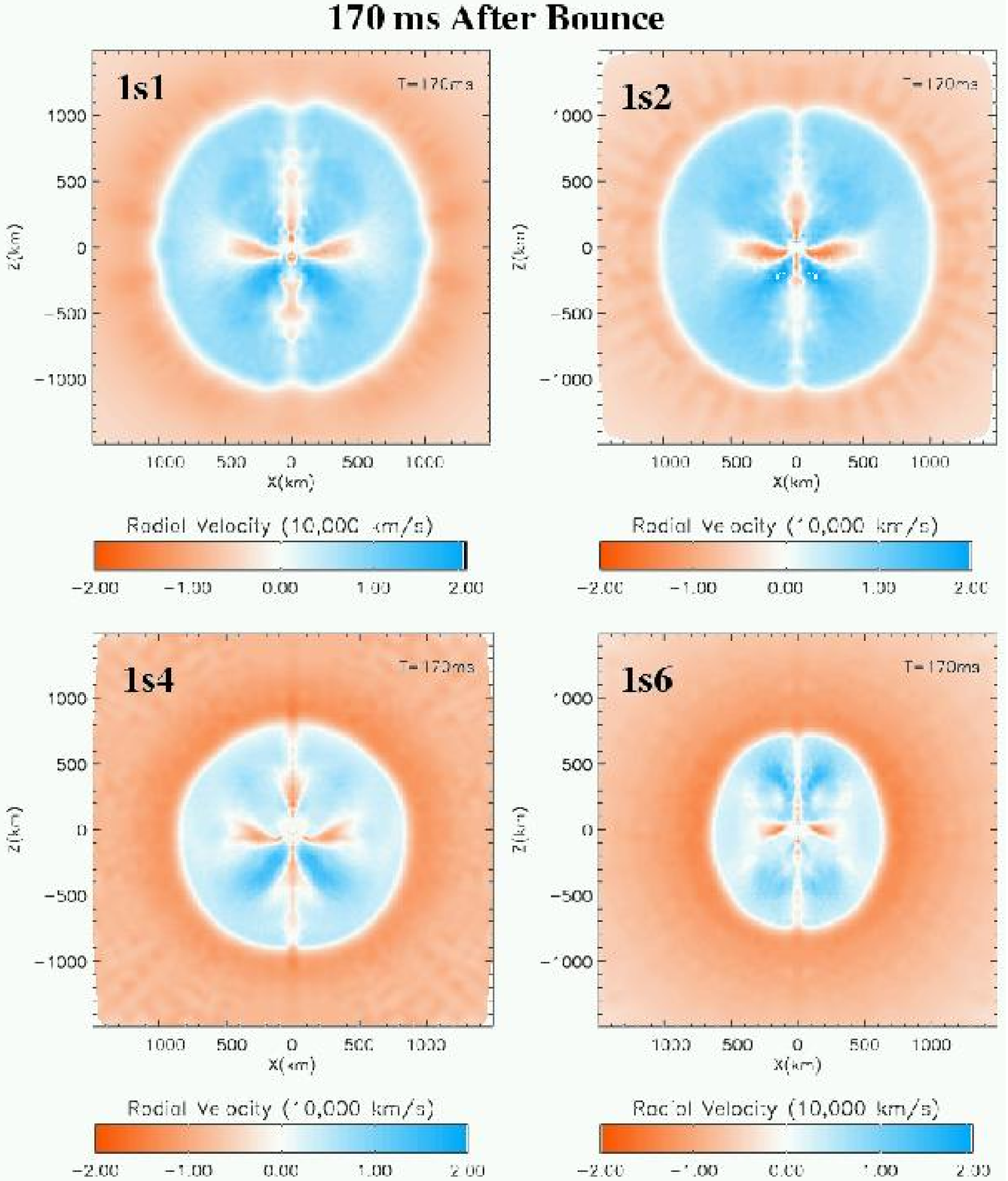}
\figcaption{\Figvs}
\vspace{1.5\baselineskip}}{}

A few other characteristics of the collapse also affect the explosion.
The neutrino luminosity from the proto-neutron star core
determines how much energy is deposited in the convective region and,
ultimately, the explosion energy (again, see Burrows \& Goshy 1993,
Janka 2000).  In general, as the neutrino emission from the core
increases, the chance that an explosion will occur also increases.
Figure 9 shows the electron-neutrino luminosity for our four
comparison models.  Note that although the neutrino luminosities for
Models 1s1, 1s2, and 1s6 are all nearly equal, Model 1s4 emits nearly
twice as many electron neutrinos 100\,ms after bounce.  The large iron
core of Model 1s4 compresses much less during collapse than the cores
of the other stars, and a larger fraction of its energy is emitted via
electron neutrinos instead of $\mu$ and $\tau$ neutrinos.  Electron
neutrinos deposit energy into the convective region much more
efficiently than $\mu$ and $\tau$ and this helps to explain the
fact that model 1s4 eventually explodes and Model 1s6 does not.

\pFig{nu}
\ifthenelse{\boolean{emul}}{
\vspace{1.5\baselineskip}
\noindent
\includegraphics[angle=0,width=\columnwidth]{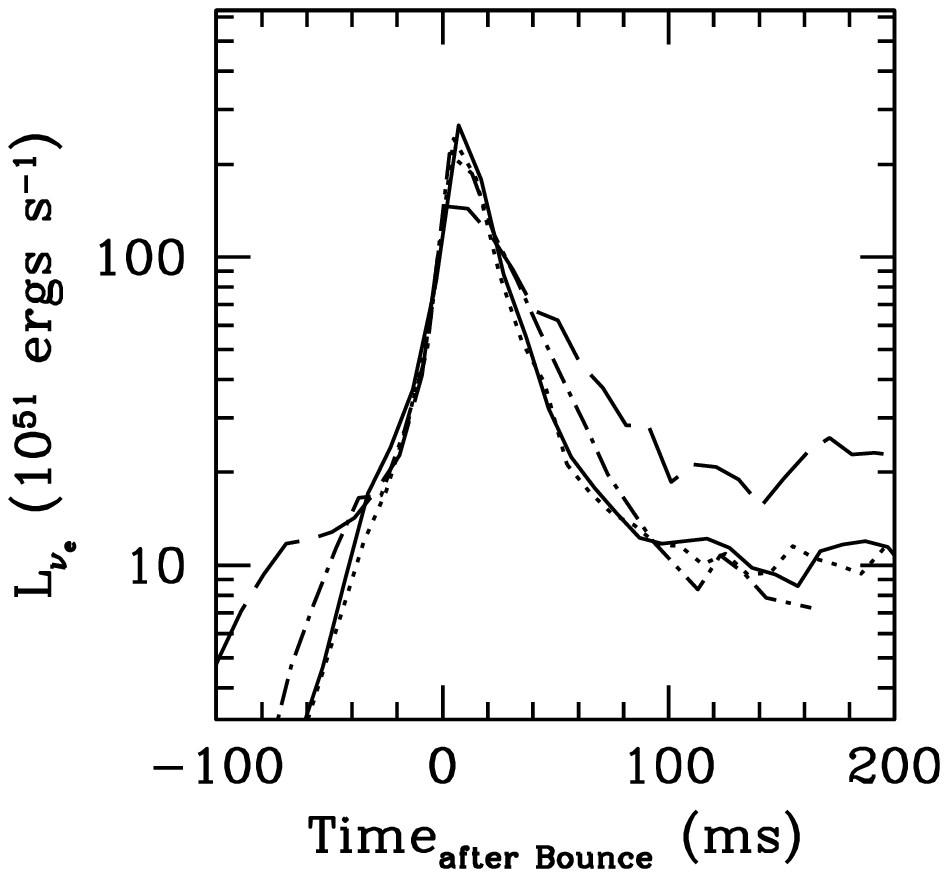}
\figcaption{\Fignu}
\vspace{1.5\baselineskip}}{}

The efficiency of neutrino energy deposition can also differ slightly
from model to model.  The entropy profile left behind when the bounce
shock stall differs from star to star (Fig.\ 10).  It seeds
convection, which enhances the efficiency of neutrino deposition. 
Note that the entropy profile set by the bounce shock is much higher
in Models 1s1 and 1s2 than in Models 1s4 and 1s6, producing stronger
initial convection.

\pFig{sf}
\ifthenelse{\boolean{emul}}{
\vspace{1.5\baselineskip}
\noindent
\includegraphics[angle=0,width=\columnwidth]{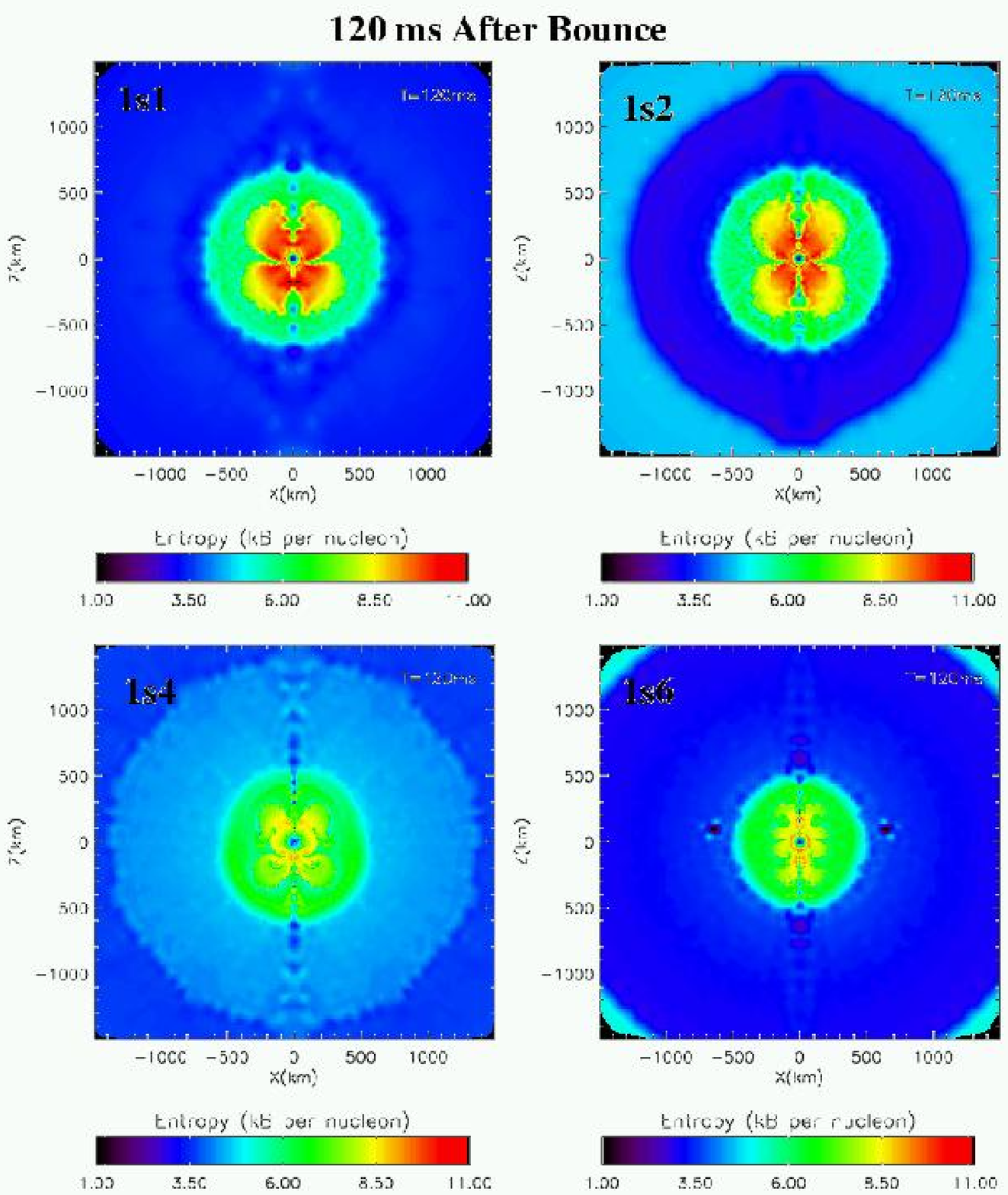}
\figcaption{\Figsf}
\vspace{1.5\baselineskip}}{}

\subsection{Fallback and Remnant Masses}

To calculate the remnant masses, we must follow the supernova explosion
to late times ($\sim 1$\,yr), and to do this, we map our 2-dimensional
results back into 1-dimension.  This simplification is necessary to
make the calculation computationally tractable and since the shock
is roughly symmetric, this mapping can be accomplished smoothly.
In addition, the outer edges of our 2-dimensional simulation do
not move much over the course of the $\sim 0.5$\,s collapse
simulation and adding the rest of the star (in the models
1s3, 1s4, and 1s6) is also straightforward.  Using a 1-dimensional
core-collapse code (Herant et al. 1994), we can then follow the
explosion to later times.

Even in 1-dimension, we must make some approximations to speed up the
code.  First, we remove the proto-neutron star and place an inner
boundary at its surface.  In addition, as material falls back onto the
proto-neutron star and piles up onto its surface, we accrete this
material onto the neutron star.  This is achieved numerically by
removing a zone when its density rises above a critical density and
adding its mass to the proto-neutron star.  If the critical density is
too low, material can artificially accrete too rapidly onto the
neutron star and the value of our critical density
($10^{11}$\,g\,cm$^{-3}$) was chosen to avoid this artificial
accretion.  By running the explosion out to 1\,yr after collapse, we
calculate the final remnant mass (baryonic) and the final kinetic
energy of the explosion (Table 2).
Because of the energy required to unbind the envelope, the final kinetic
energy is much less than the initial explosion energy.
Depending on the modest variations in the Wolf-Rayet mass loss rate
(a factor of~3), the death of a 60\,\Msun star in a close binary system
can produce anything from a low mass nutron star to a 10\,\Msun black hole.

Figure 11 shows the mass trajectories of model 1s4, which has the most
supernova fallback after an initial successful explosion (remember
that Model 1s6 did not successfully launch an outward moving shock).
In this simulation, it took nearly a year for all of the fallback
material to ultimately accrete onto the neutron star.

\pFig{fall}
\ifthenelse{\boolean{emul}}{
\vspace{1.5\baselineskip}
\noindent
\includegraphics[angle=0,width=\columnwidth]{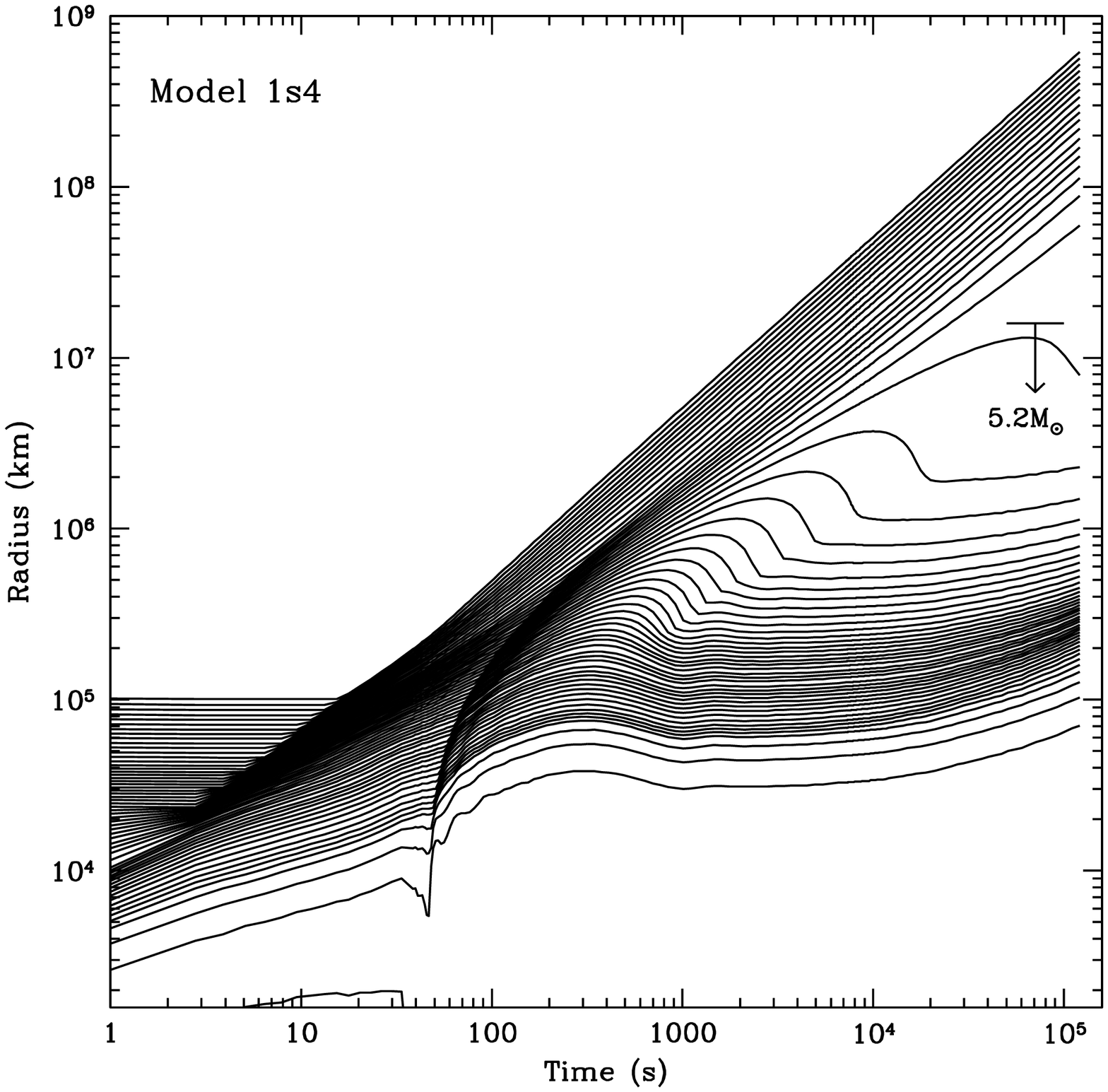}
\figcaption{\Figfall}
\vspace{1.5\baselineskip}}{}

\section{Summary}

In this paper, we present models for
life and death of a series of 60\,\Msun stars in
binaries, whose evolution differs only by the mass loss rate adopted
in the Wolf-Rayet stage.  The Wolf-Rayet mass loss rate is still very
uncertain and has only recently been revised downward by a factor
2$\ldots$3 (Hamann \& Koesterke 1998).  In addition to being uncertain,
it is likely that the Wolf-Rayet mass-loss rate depends on metallicity.
Aside from the mass-loss rate, the metallicity has very little effect
on the stellar evolution models.  Thus, the study presented
here can also be regarded as a study of this binary at different
metallicities -- the uncertainty of the WR mass loss rate then
translates into a variation of the
the initial stellar metal abundance.

We obtain final pre-collapse stellar masses in the range from
3.1\,\Msun for the highest mass loss rate (Model~1s1) up to
10.7\,\Msun for the lowest mass loss rate (1s6), while the central
carbon abundance at core helium exhaustion drops form 35\,\% to
22\,\%, respectively.  The ensuing complex interaction of carbon core
and shell burning phases with later burning stages causes a
non-monotonic behavior of the pre-collapse structure of the star,
i.e., the masses of the neon-oxygen core, the silicon core or the
deleptonized core (\Tab{evo}; \Sect{Cc2cc}).

Following the subsequent collapse and supernova explosion of these
stars, we find that these models produce a range of compact
remnants from a 1.17\,\Msun neutron star (1s2) to a 10.7\,\Msun black
hole (1s6).  The remnant mass does not scale strictly with the iron
core mass (it depends on the density and temperature structure of the
collapsing core - see \S 3) and we can not use this core mass to
estimate the supernova explosion energy or the compact remnant mass.
Amazingly, the large differences in the remnant mass are caused by
only a factor of~ 3 change in the Wolf-Rayet mass-loss rate, and a
40\% difference in the amount of mass lost through Wolf-Rayet winds.
In view of this extreme sensitivity of the remnant mass on the WR mass
loss, the persisting uncertainty of the WR mass loss rate, and the
uncertainties of our core collapse models, it is difficult to draw
solid conclusion.

Nevertheless, taking our results at face value implies that to form a
black hole of 10\,\Msun or more from a 60\,\Msun star in a Case~B (or
Case~A) binary, one might need to use WR mass loss rates smaller than
the currently favoured ones (unless the system is at lower
metallicity).  That is, it seems difficult (though not
impossible) to form those X-ray binaries which contain the most
massive black holes, like Cyg~X1 or V404~Cyg, through Case~A or~B at
solar metallicity, and the Case~C scenario (Brown et al. 1999, 2001,
Kalogera 2001, Fryer \& Kalogera 2001) may provide a viable
alternative.  For most low mass black hole binaries, which may contain
black holes with 3$\ldots$7\,\Msun (Fryer \& Kalogera 2001), our
results imply that a Case~B progenitor evolution may be sufficient if
we assumed that the mass loss rate were reduced by a factor of
$\sim$4.  Given the uncertainties in WR mass loss, stellar evolution,
and core-collapse, Case~B progenitors for most black hole binaries are
not excluded.

Furthermore, only with our lowest mass loss rate were we able to produce 
a direct collapse black hole from this Case~B progenitor.  
Fryer, Woosley, \& Hartmann (1999) suggested that most collapsar
$\gamma$-ray burst progenitors are produced in binares that undergo Case~B 
mass transfer, but our results imply
that such binaries may not produce collapsars (at least at solar 
metallicity).  However, most $\gamma$-ray bursts occur at high 
redshifts and low metallicities.  That is, if the Wolf-Rayet 
mass loss rates truly decrease with decreasing metallicity, 
this does not preclude Case~B progenitors of $\gamma$-ray bursts.

All in all, our results should be understood as temporary, awaiting a
better understanding of the Wolf-Rayet winds and the core collapse of
massive stars.  But to understand the origin of the black hole
binaries in our nearby universe, we must continue to pursue Case AB,
as well as Case~C models.  We should finally mention that we left out
two potentially important stellar parameter which is essential for
most current models of collapsing stars: rotation and magnetic fields.
Rotation may add another dimension to the expected remnant mass as
function of stellar parameters, which has to be left here for future
investigations.  Although Fryer \& Heger (2000) found that the
currently proposed dynamos would not develop strong enough magnetic 
fields to drive the explosion alone, magnetic fields may still effect the 
explosion and should be considered.

\acknowledgements This research has been supported by NASA (NAG5-2843,
MIT SC A292701, and NAG5-8128), the NSF (AST-97-31569), the US DOE
ASCI Program (W-7405-ENG-48), a LANL Feynman Fellowship, the
Alexander von Humboldt-Stiftung (FLF-1065004), and the Deutsche
Forschungsgemeinschaft (La 587/15-2).

{}

\clearpage

\begin{deluxetable}{lrrrrrrrrrrr}
\tablewidth{40pc}
\tablecaption{Model Parameters and Stellar Evolution Results\lTab{evo}}
\tablehead{ \colhead{mod.} &
\colhead{$\fWR$\tablenotemark{a}} &
\colhead{$\tWNL$\tablenotemark{b}} &
\colhead{$\tWNE$\tablenotemark{b}} &
\colhead{$\tWCO$\tablenotemark{b}} &
\colhead{$\Delta M_{\rm WR}$\tablenotemark{x}} &
\colhead{$\Cc$\tablenotemark{c}} &
\colhead{$\MHe$\tablenotemark{d}} &
\colhead{$\MCO$\tablenotemark{d}} &
\colhead{$\MNeO$\tablenotemark{d}} &
\colhead{$\MSi$\tablenotemark{d}} &
\colhead{$\MYe$\tablenotemark{d}} \\
\colhead{} & \colhead{} & \colhead{(\kyr)} & \colhead{(\kyr)} &
\colhead{(\kyr)} & \colhead{(\%)}& \colhead{(\Msun)}&
\colhead{(\Msun)}& \colhead{(\Msun)}& \colhead{(\Msun)} &
\colhead{(\Msun)} }

\startdata

1s1 & 1/1 &  25   &  24  & 776 & 23.7 & 35 &  3.132 & 2.373 & 1.672 & 1.566 & 1.352 \\
1s2 & 1/2 &  55   &  41  & 494 & 22.4 & 30 &  4.389 & 3.375 & 1.849 & 1.774 & 1.305 \\
1s3 & 1/3 &  73   &  41  & 373 & 20.7 & 27 &  6.108 & 4.782 & 2.514 & 2.029 & 1.606 \\
1s4 & 1/4 &  85   &  44  & 318 & 19.2 & 25 &  7.550 & 5.934 & 2.934 & 2.308 & 1.749 \\
1s6 & 1/6 &  122  &  53  & 205 & 16.1 & 22 & 10.746 & 8.545 & 2.330 & 1.939 & 1.497 \\

\enddata

\tablenotetext{a}{Wolf-Rayet mass loss relative to Braun (1997)}

\tablenotetext{b}{lifetime of the star in the WNL (\tWNL),
the WNE (\tWNL), and WC+WO (\tWCO) phases}

\tablenotetext{c}{central carbon mass fraction after central helium
depletion}

\tablenotetext{x}{total amount of mass lost during the Wolf-Rayet stage}

\tablenotetext{d}{mass of the (helium) star (\MHe), the
helium-free carbon-oxygen core (\MCO), the carbon-free
neon-magnesium-oxygen core (\MNeO), the oxygen-free silicon
core (\MSi), and that of the deleptonized core (\MYe;
defined by $\Ye<0.497$)}

\lTab{param}
\end{deluxetable}
\clearpage

\begin{deluxetable}{lccccc}
\tablewidth{26pc}
\tablecaption{Results from the core collapse calculations}
\tablehead{ \colhead{Model\tablenotemark{a}} &
\colhead{$M_{\rm Fe Core}$\tablenotemark{b}} &
\colhead{$E_{\rm exp}$\tablenotemark{c}} &
\colhead{$t_{\rm exp}$\tablenotemark{d}}
& \colhead{$KE_{\infty}$\tablenotemark{e}}
& \colhead{$M_{\rm Remnant}$\tablenotemark{f}} \\
\colhead{} & \colhead{(\Msun)} & \colhead{($10^{51}$ergs)}
& \colhead{(ms)} & \colhead{($10^{51}$ergs)} & \colhead{(\Msun)}}

\startdata

1s1  & 1.321 & 1.45 & 150 & 0.7 & 1.35 \\
1s2  & 1.352 & 2.36 & 160 & 1.3 & 1.17 \\
1s3  & 1.590 & 1.60 & 180 & 1.0 & 2.11 \\
1s4  & 1.750 & 0.3  & 180 & 0.15 & 5.2 \\
1s6  & 1.497 &  0   &  -  &  0  & 10.7 \\

\enddata

\tablenotetext{a}{see \Tab{param}}
\tablenotetext{b}{Mass of the iron core at collapse.}
\tablenotetext{c}{Amount of energy injected in the star by neutrinos.
A portion of these energy will unbind the star, a portion will go into
the velocity of the ejecta, and some will fall back onto the neutron
star.}
\tablenotetext{d}{Time it takes for the shock to be pushed beyond
1000\,km.}
\tablenotetext{e}{Kinetic energy of the ejecta 1 year after explosion.}
\tablenotetext{f}{Mass of the compact remnant after fallback.}

\lTab{60msun}
\end{deluxetable}
\clearpage

\ifthenelse{\boolean{emul}}{}{

\clearpage
\onecolumn

\ifthenelse{\boolean{\IncludeFigures}}{
\renewcommand{\figcaption}[2][]{
\clearpage
\begin{figure}
\plotone{#1}
\caption{#2}
\end{figure}
}}{}

\epsscale{0.8}\figcaption[\FighecoreFile]{\Fighecore}
\epsscale{0.75}\figcaption[\FigconvOneFile]{\FigconvOne}
\epsscale{0.75}\figcaption[\FigconvTwoFile]{\FigconvTwo}
\epsscale{0.75}\figcaption[\FigconvFourFile]{\FigconvFour}
\epsscale{0.75}\figcaption[\FigconvSixFile]{\FigconvSix}
\epsscale{1.0}\figcaption[\FigmdotFile]{\Figmdot}
\epsscale{0.8}\figcaption[\FigvfFile]{\Figvf}
\epsscale{0.8}\figcaption[\FigvsFile]{\Figvs}
\epsscale{1.0}\figcaption[\FignuFile]{\Fignu}
\epsscale{0.8}\figcaption[\FigsfFile]{\Figsf}
\epsscale{0.8}\figcaption[\FigfallFile]{\Figfall}
}

\end{document}